\documentclass[aps,prb]{revtex4}
\usepackage{bm}
\usepackage{amsmath}
\usepackage{color}
\usepackage{graphicx}
\usepackage{amssymb}
\DeclareMathAlphabet{\mathbbmsl}{U}{bbm}{m}{sl}
\makeatletter
\newsavebox{\@brx}
\newcommand{\llangle}[1][]{\savebox{\@brx}{\(\m@th{#1\langle}\)}%
    \mathopen{\copy\@brx\kern-0.5\wd\@brx\usebox{\@brx}}}
\newcommand{\rrangle}[1][]{\savebox{\@brx}{\(\m@th{#1\rangle}\)}%
    \mathclose{\copy\@brx\kern-0.5\wd\@brx\usebox{\@brx}}}
\makeatother

\begin{document}
\draft

 \title{Anisotropic optical conductivities of Model Topological nodal-line  Semimetals}

\author{Sita Kandel$^{1,2}$, Godfrey Gumbs$^{1,2,3}$,
and Oleg L. Berman$^{2,4}$  }
\address{$^1$Department of Physics, Hunter College, City University of New York, 695 Park Avenue, New York, NY 10065 USA}
\address{$^{2}$The Graduate School and University Center, The
City University of New York,    New York, NY 10016, USA}
\address{$^3$Donostia International Physics Center (DIPC), P de Manuel Lardizabal, 4, 20018 San Sebastian, Basque Country, Spain}
\address{$^{4}$Physics Department, New York City College
of Technology, The City University of New York \\
Brooklyn, NY 11201  USA}

\date{\today}

\begin{abstract}

With the use of simple models, we investigated the optical conductivity of a nodal-line semimetal  (NLSM) whose crossing of the conduction and valence bands near the origin ($O$ point) in the ($k_x,k_y$) plane of a small  cubic region can be adjusted by a parameter $\alpha$.   The Hamiltonian of the NLSM is based on the ${\bf k}\cdot {\bf p}$  model for the low-lying energy bands.   When $\alpha=0$, these bands touch each other along a continuous closed loop but the  opening of a band gap corresponding to finite values of $\alpha$ and the varying of the carrier concentration can be adjusted.   This provides a tunable semiconductor  gap, around the $O$ point and the valence and conduction bands   can meet  at a pair of points within the small cubic region in ${\bf k  }$  space.   The optical conductivity of such a NLSM is calculated using the Kubo formula with emphasis on the optical  spectral weight  redistribution, deduced from appropriate Green's functions, brought about by changes in gap and chemical potential due to modifying   $\alpha$. We  derived closed-form semi-analytic  expressions for the longitudinal components of the optical conductivity for these model systems of NLSM and compare results for chosen $\alpha$ and chemical potential. We also present results for the heat capacity when the system is in thermal equilibrium for various chosen $\alpha$ and chemical potential.\end{abstract}

\maketitle
\medskip

\noindent
{Corresponding Author}:\ \  Oleg L. Berman, Email:  OBerman@citytech.cuny.edu.

\medskip
\section{Introduction}
\label{sec1}

 Topological semimetals (TSM) have been emerging as a fast growing group of  topological  materials which have been coming under increasing scrutiny.\cite{PRX, ZrSiSe, NaturePhys}   For these systems, the conduction and valence bands cross each other in the Brillouin zone.  It is important to note that the crossing cannot be eliminated by perturbing the Hamiltonian without breaking either its crystalline  or time-reversal symmetry. For Dirac semimetals, the conduction and valence bands have linear crossing at the Dirac point. If the time reversal symmetry is broken, the Dirac point splits into two separate ones, and the system becomes a Weyl semimetal. On the other hand, for the nodal-line semimetal (NLSM), the crossing of the valence and conduction bands forms a nodal ring (also called nodal line). There is a significant difference in the density-of-states near the Fermi surface between the NLSM compared to Dirac and Weyl semimetals. Due to the linear energy dispersion, the density-of-states of Dirac and Weyl semimetals is linear in energy and vanishes at the Fermi level. We do not find this behavior in our model of NLSM.  For 3D TSM, two bands may cross one another at discrete points or along a closed curve.

\medskip
\par

For a TSM, one can associate for each band crossing a topological invariant which depends on the symmetry group that protects the nodal structure. Various material systems have been proposed as NLSM protected by different symmetry groups. For example, TaAs \cite{belopolski} has been shown to  be a NLSM protected by mirror reflection and spin rotation symmetries with two nodal lines in the absence of spin orbital coupling (SOC). However, in the presence of SOC,  each nodal line is grouped into three pairs of Weyl nodes. SrIrO$_3$ \cite{junwei} is another NLSM with a double nodal line. This material is separated into a pair of Dirac nodes under broken mirror reflection symmetry.

\medskip
\par

For many years, the topological phases of matter, topological insulators and topological semimetals, both Dirac and Weyl semimetals,  have been widely studied  theoretically and experimentally. When it came to be known that some topological semimetals exhibit different dispersion relations than that of Dirac and Weyl, making a nodal ring at the cross-section of valence and conduction bands, NLSM have attracted tremendous interest in condensed matter physics. The optical conductivity of NLSM has also been studied both theoretically and experimentally \cite{carbotte, mukherje, barati, kohino, 1PRBRef2, wang, strain, Qiu,Ebad-Allah1, Ebad-Allah2, Ebad-Allah3}. For example, in 2016/17,  Carbotte and his team \cite{carbotte, mukherje} studied the optical response of such NLSM  as a function of different parameters using the Kubo formula. They presented analytical and numerical solutions for the optical conductivity of a NLSM using a toy model Hamiltonian. They came to the conclusion that their 3D NLSM presents 2D Dirac-like response in the low-photon energy regime and 3D Dirac-like response in the high-energy photon limit, i.e. the optical conductivity is increased linearly   with photon frequency. Later on in 2017,   Barati \cite{barati}also showed that the optical conductivity of NLSM is anisotropic.The response along the direction parallel to the plane containing the   nodal ring increases linearly while that along the direction perpendicular to the nodal ring saturates to constant value at high frequencies. Both those authors considered different toy models in order  to explain NLSM with different Fermi surface. However, neither of these results totally agree with the paper by  Habe and Koshino \cite{kohino} in 2018. Koshino and his team studied theoretically the dynamical conductivity of NLSM ZrSiS by using a multi-orbital tight-binding model based on a first-principles energy  band calculation. According to Habe and Koshino, for ZrSiS type NLSM, the interband contribution first increases slowly for some frequency range and is decreased  to smaller value and saturates for large frequency.

\medskip
\par
More advanced and very sophisticated model Hamiltonians have been used to capture the behavior of properties of materials with line nodes. For instance, the authors in Ref.\  [\onlinecite{1PRBRef2}] have considered a heterostructure  which is in a topological  semimetallic phase with line nodes. By  employing the Kubo formula, they have investigated the dependence of the dc conductivity on magnetization $M$ for a structure consisting of alternate layers of topological insulators where spin splitting, resulting from magnetization in the z direction was  included.  Also tunneling within a topological insulator layer and between neighboring topological insulator layers was also taken into consideration  and explicitly contribute to the model Hamiltonian that displays a number of anomalous properties and possess novel correlated phases. The dependence of the dispersion on the strength of the time-reversal breaking plays a crucial role in this model. Likewise, the paper by Yan, et al.\cite{2PRBRef2} deals with the collective modes of nodal line semimetals.  With their model Hamiltonian, they were able to obtain the plasmon dispersion relation which in the regime of modest doping and in the long wavelength limit, the frequency  of the mode was shown to obey $\omega_p\sim  n^{1/4}$ where $n$ is the carrier density. This is in contrast with ordinary electron liquids and the Dirac-Weyl liquids.  But, in the regime of very low doping the $\omega_p\sim  n^{1/4}$  law crosses over to $\omega_p\sim  n^{1/2}$.  These calculations were done in the random-phase approximation with the polarization function determined using a bubble consisting of single-particle Green's functions  without self-energy corrections.

\medskip
\par
 This brief overview has motivated us to study the optical response of materials similar to  Dirac semimetal and Weyl semimetal using a simple model Hamiltonian in conjunction with the Kubo formula which yields semi-analytic results for the optical conductivity.  Our results  include contributions from both intra and interband transitions between the low-energy subbands for our model Hamiltonian. The obtained results are interesting because the numerical solutions  have similar features to Habe and Koshino's results \cite{kohino} and are neither like 2D nor 3D Dirac or Weyl semimetal. The transverse optical conductivity of NLSM is found to vanish due to rotational symmetry along the ring of NLSM. This result is also supported by  a recent study of  Wang \cite{wang}, who demonstrated the anomalous Hall optical conductivity in tilted topological NLSM. We refer to Ref. [\onlinecite{ChiJou}]    for a review of other remarkable properties of these TSM systems as well as a discussion of possible technological applications. The optical conductivity of monolayer and bilayer graphene, and few-layer epitaxial  graphite has been reported recently\cite{nicol}. These studies have yielded useful information regarding the electron dynamics and this is a plausible reason for our present investigation into NLSM. Here, we consider the optical conductivity of a NLSM model system with special emphasis on the optical spectral weight redistribution  due to changes in the chemical potential caused by charging  as well as an adjustable parameter $\alpha$ which governs the degree of crossing or the gap between the valence and conduction bands. The doping determines the position of the Fermi level which could be located within the energy gap or the conduction sub band. The parameter $\alpha$ can be manipulated by strain and the doping could be varied through the intercalation of donor atoms or by applying a voltage to metal gate. The effect of mechanical strain on the optical properties of NLSM ZrSiS has been studied using first-principles calculations in Ref. [\onlinecite{strain}]. According to Ref. [\onlinecite{strain}], frequency-independent optical conductivity is robust with respect to uniaxial compressive strain of up to 10 GPa.   Therefore, our reported studies related to the possible control of the optical conductivity of a model NLSM system by external strain is likely to attract attention from  theoretical and experimental researchers.

\medskip
\par
It is also worthy of note that for slabs of finite thickness unusual surface states, Fermi-arc surface states in Weyl semimetal \,\cite{ S2015Xu, NP2016Deng}, and drumhead surface states  in NLSM \cite{DH2020Hosen, PRX, DH2016Bian, DH2020Wang, DH2017Li, DH2015Weng} are supported by TSM. So far, only bulk and surface Landau levels, together with their transport properties, have been extensively investigated for Dirac semimetal \cite{NP2014Fu, PRL2011Apa, PRB2010Hana, PRL2010Cheng, ACS2020Chong} and Weyl semimetal \cite{NC2018Yuan, PRB2018Say,PRB1983Nie, PRB2020Mat, SR2016Zhang, PRL2017Wang}. We emphasize that in this paper, we  consider only a bulk NLSM. However, we may model the effects due to a surface by a slab containing multi-layers of single atoms with two orbitals and  stacked one on top the other along a specific direction. The surface states of topological materials are significantly affected by the number of layers and impurities.  When the number of material layers is insufficient, the surface state will not be formed, or impurities in the system can readily destroy the surface states. In general, observation of surface states of topological materials such as  drumhead surface states of NLSM\cite{DH2020Hosen, PRX, DH2016Bian, DH2020Wang, DH2017Li, DH2015Weng} is still a great challenge.  Our current work mainly focuses on the physical properties of bulk states.

\medskip
\par

The rest of this paper is organized as follows.  In Sec.\  \ref{sec2}, we present a continuum  model Hamiltonian for low-energy  subbands based on the ${\bf k}\cdot {\bf p}$ method.   These energy bands are displayed within a small cubic region. We also present the theoretical formalism  for calculating the optical conductivity based on this simple Hamiltonian.  This model involves an adjustable parameter $\alpha$  associated with a nodal circle which may shrink to a point  when $\alpha=0$. Here, we have managed to derive a conveniently simple semi-analytical expression for the frequency-dependent conductivity for arbitrary chemical potential, making this a convenient result for experimentalists to explore further.  In Sec. \   \ref{sec3}, we present numerical results demonstrating the  anisotropy of the system, the $\alpha$-dependence and charging pertaining to the biasing of the NLSM sample.   In Sec.\ \ \ref{sec4}, we present and discuss our numerical results for the density-of-states.  Acknowledging that the Hamiltonian in Sec.\ \ref{sec2}  is not fully representative of the band structure properties of a specific NLSM but only captures  some essential properties of this class of materials, we present another model Hamiltonian in Sec.\  \ref{sec5}. We discovered that both models have similar behaviors for their optical conductivities and so may be employed  to make some important predictions about the optical properties of NLSMs.  We present in Sec. \ \ref{sec6} numerical results for the heat  capacity of a NLSM in thermal equilibrium.  The formalism is rather straightforward and makes use of the grand potential.   Section \ \ref{sec7}  is devoted to a summary of our results. Some of the algebraic calculations are presented in appendices.

\section{Theoretical Background}
\label{sec2}

\begin{figure}[ht]
\centering
\includegraphics[width=0.5\linewidth]{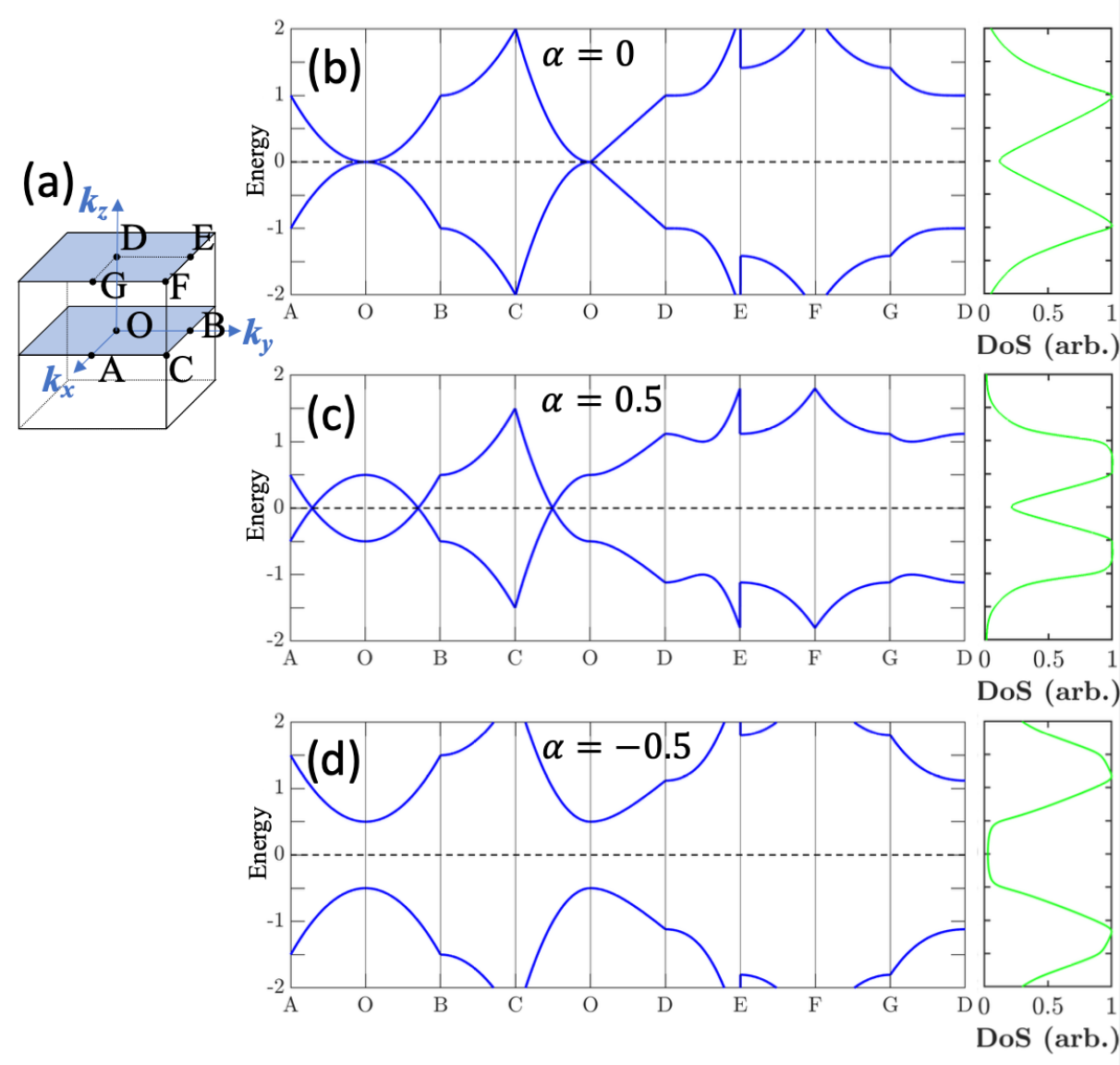}
\caption{ (Color online)  (a)   shows   the cubic region with some chosen symmetry points for the bulk NLSM. Bulk band structures between these  symmetry points (left) and corresponding density-of-states (right) for (b) $\alpha=0$, (c) $\alpha=0.5 $ and (d) $\alpha=-0.5 $ .  }
\label{energy}
\end{figure}

Let us consider a Hamiltonian for which  the
nodal line is stable against perturbations.  However,  the nodal
line may still shrink continuously to a point.  In this regard, we
turn to  a single-spin effective Hamiltonian  of  a material such as
ZrSiSe and ZrSiTe.  \cite{ZrSiSe,NaturePhys}   In Ref.\
[\onlinecite{PRX}], the band structure of ZrSiTe was obtained using
{\em ab initio\/} calculations. The idea here is to carry out a
calculation based on a model Hamiltonian which mimics the essential
features of the true band structure near the $\Gamma$ point of a
NLSM such as ZrSiTe. In general, these band structures obtained from
first-principles calculations or the generalized tight-binding model
are very complicated. Of course, not all features are reproduced by
this model Hamiltonian throughout a small  cubic region to which the model is restricted  but it could
serve as a useful tool for generating qualitative results to get a
better understanding of this growing class of topological
semimetals. For the nodal line, protected by inversion, time-reversal and spin rotation $SU(2)$ symmetries, the symmetry group $Z_{2}$ is the topological classification
of the wave functions on the ring/sphere.~\cite{ChiJou} Therefore, a
nodal ring is characterized by two independent $Z_{2}$ indices,
denoted by $\zeta_{1}$ and $\zeta_{2}$, defined on a ring that links
with the line and on a sphere that encloses the whole line.
According to Ref.~[\onlinecite{ChiJou}], all topological nodal
rings, protected by this symmetry group with respect to an
arbitrarily small perturbation, are characterized by $\zeta_{1} =
1$. For  NLSM with $\zeta_{1} = 1$ and $\zeta_{2} = 0$, our toy
model Hamiltonian is given by \cite{ChiJou}

\begin{equation}
\hat{H}=  (\alpha-k^2)\hat{\sigma}_z + k_z\hat{\sigma}_x  \  ,
\label{Hamil1}
\end{equation}
where energy is measured in units of $\hbar v_F k_c$ and the wave vector ${\bf k}=(k_x,k_y,k_z)$ is measured in units of $k_c$ with the velocity  $v_F \sim 1.5\times 10^5$m/s and $k_c  $ is the wave vector corresponds to momentum cut-off.  Also, the energy is scaled with respect to $\hbar v_F k_c$, all wave vectors are scaled with respect to  $k_c$ and  $-1\leq k_i\leq 1$ with $i=x,y,z$ in order to ensure that the model is restricted near the $O$ point.  For convenience, we refer to this region as the cubic region.  We  have $\hat{\sigma}_x,\hat{\sigma}_z$ representing Pauli matrices. The quantity $\alpha$ is an adjustable parameter which can be employed to vary the energy gap around the $O$ point for which the energy bands are given by

\begin{equation}
\epsilon_{s}({\bf k}) =s \sqrt{k_z^2  +(\alpha-k^2)^2} \   .  
\label{eigenvalue}
\end{equation}
These bands touch at ${\bf k}=0$  when $\alpha=0$. If $\alpha>0$,
the two subbands  cross each other  near the $O$ point on the
$ (k_x,k_y) $ plane for fixed $k_z=0$, describing a nodal circle of
radius $\sqrt{\alpha}$  which shrinks to a point at ${\bf k}=0$ as
$\alpha $ is decreased to zero and  the nodal circle vanishes when
$\alpha<0$.  In Fig.\  \ref{energy}, we plot the energy bands in the
small cubic region along lines joining given points, corresponding
to $\alpha=0$ and $\alpha=0.5 $. These two contrasting cases
demonstrate how the band gap near $O$ may be tuned by varying
this parameter  which may be manipulated as a function of strain. There are several mechanisms which one may use to engineer and tune the band structure. These include the  interactions with substrates \cite{Kek3}, strain \cite{Kek4,Kek5,Kek6} , and time-dependent electromagnetic fields \cite{Kek14,Kek15} .  It is clear from these results that the band
structure is modified by finite $\alpha$  throughout the cubic region. Figure\
\ref{energy}(b)  shows that the bulk band  gap closes at an even
number of  discrete  points near the $O$ point.  These special
gap closing points in the cubic region are protected by
crystalline symmetry and play a role in the  behavior of Weyl
semimetals.\cite{Nag} The detailed analysis of the topology of
NLSMs, corresponding to the Hamiltonian in Eq.\ (\ref{Hamil1}), is
provided in Ref.~[\onlinecite{ChiJou}]. This modification of the
band structure has significant effects on the physical properties we
calculate in this paper and may also affect the thermoelectric and
Boltzmann transport as well as the plasmon excitations. In our model Hamiltonian, the gap is controlled by a parameter $\alpha$ as shown in Eq. (1) in the manuscript. 

\begin{figure}[ht]
\centering
\includegraphics[width=0.5\linewidth]{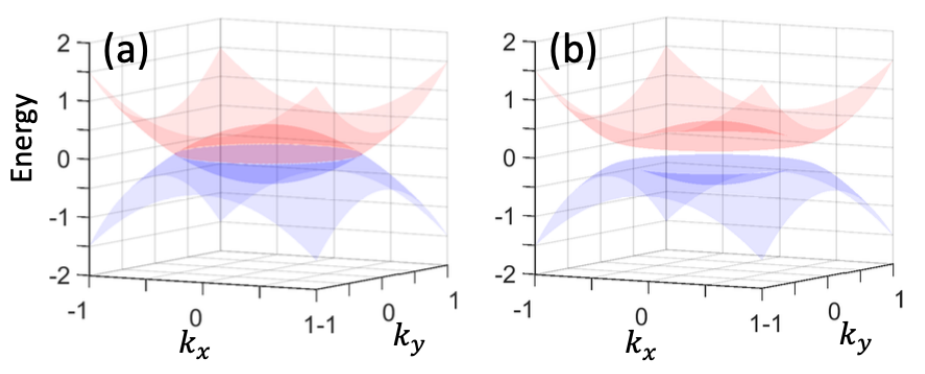}
\caption{ (Color online) The 3D contour bands at low energy on the selected  ($k_x, k_y$) plane  for (a) $k_z=0$ and (b)  $k_z=0.2$ at $\alpha=0.5$. The pink curve corresponds to the conduction band ($s=+$) and the blue curve corresponds to the valence band ($s= -$ ) }
\label{energy2}
\end{figure}


\medskip
\par
 The normalized eigenvectors $\varphi_{s}$ (for $s=\pm$) of the Hamiltonian  (\ref{Hamil1}) are 
 \(
\varphi^T_{s}({\bf k}) = \left(
 \beta_{s} ({\bf k})\  ,  \eta_{s}({\bf k})
 \right)  \frac{e^{i{\bf k}\cdot{\bf r}}}{\sqrt{\cal V}}
\).      
 
where ${\cal V}$ is  a normalization volume
and $\beta_{s}$ and $\eta_{s}$ are given by
$\beta_{s} ^{2}({\bf k}) = \frac{k_{z}^{2}}{k_{z}^{2} + \left(k^{2} - \alpha \  +s
\sqrt{\left(k^{2} - \alpha\right)^{2} + k_{z}^{2}}\right)^{2}} \  $ and 
$\eta_{s}^{2}({\bf k}) = \frac{\left(k^{2} - \alpha \ +s\sqrt{\left(k^{2} - \alpha
\right)^{2} + k_{z}^{2}}\right)^{2}}{k_{z}^{2} + \left(k^{2} - \alpha \ +s
\sqrt{\left(k^{2} - \alpha \right)^{2} + k_{z}^{2}}\right)^{2}} $ with 
 $\beta_{s}^{2} +\eta_{s}^{2} = 1$.

\medskip
\par
Figure  \ref{energy}  shows the low-energy bands of the bulk NLSM along the line joining the symmetry points in the cubic region are anisotropic along  the three perpendicular axes. When we go from $O$ to $A$ and $B$, the energy bands are symmetric but it is not so along the $OD$ axis.  For $\alpha=0$, the two subbands touch at $O$ making a flat dispersionless band in a small region around it. It is clearly seen that the dispersion around $O$ is not linear like that for Dirac or Weyl semimetal. Therefore, we should not expect Dirac or Weyl-like optical conductivity even though two bands just touch at a point when the value of $\alpha$ is set to zero. For the numerical calculations, we have considered only two low energy bands and hence the effect of higher energy bands is neglected. The density-of- states has peak values for some energy range. The conductivity is expected to be a minimum for energy greater than that range. The 3D contour bands at low energy on the $k_x$,$k_y$ plane are either crossed or gapped depending on the value of $k_z$ as shown in Fig. \ref{energy2}. When $k_z=0$,  two bands cross. However, they open a gap on another plane for $k_z=0.2$. While integrating over the cubic region, the effects from both planes are included which leads to finite values for both interband and intraband conductivity even for pristine NLSM.

\medskip
\par
We now introduce the Green's function defined by $\hat{{\cal G}}^{-1}(z)=z\hat{I}-\hat{H}$ which in matrix form is written as

\begin{equation}
\hat{{\cal G}}(\omega , {\bf k})=\frac{1}{{\cal  D}(\omega , {\bf k})}\left( \begin{array}{cc}
\omega +(\alpha-k^2 )   & k_z\\
\\
k_z & \omega -(\alpha-k^2 )
\end{array}\right)\ ,
\label{e1}
\end{equation}
with  ${\cal  D}(\omega , {\bf k}) \equiv  \omega^2 -(\alpha -k^2)^2-k_z^2=(\omega-\epsilon_{+}({\bf k})
 )(\omega- \epsilon_{-}({\bf k}) )$.

Now, the spectral function representation of the Green's function is for $i,j=1,2$

\begin{equation}
{\cal G}_{ij}(z)=\int_{-\infty}^\infty  \frac{d\omega^\prime}{2\pi}
\frac{A_{ij}(\omega^\prime)}{z-\omega^\prime} \  .
\label{grnfn}
\end{equation}
      The elements for the spectral function defined by  Eq. (\ref{grnfn}) are 

\begin{eqnarray}
A_{11}({\bf k},\omega) &=& \frac{2\pi }{\epsilon_+-\epsilon_-} \left\{ (\epsilon_++(\alpha-k^2)) \delta(\omega-\epsilon_+) -
(\epsilon_-+(\alpha-k^2)) \delta(\omega-\epsilon_-)
\right\}
\nonumber\\
A_{22}({\bf k} ,\omega) &=& \frac{2\pi }{\epsilon_+-\epsilon_-} \left\{ (\epsilon_+-(\alpha-k^2)) \delta(\omega-\epsilon_+) -
(\epsilon_- -(\alpha-k^2)) \delta(\omega-\epsilon_-)
\right\}
\nonumber\\
A_{12}({\bf k} , \omega) &=& A_{21}({\bf k} , \omega)= \frac{2\pi  k_z}{\epsilon_+-\epsilon_-} \left\{
\delta(\omega - \epsilon_+) -\delta(\omega - \epsilon_-)\right\} \  .
\label{spectralfn}
\end{eqnarray}
An example of computing the elements of spectral function from Eq. (\ref{grnfn}) is given in the Appendix A. We can write the real part of the optical conductivity tensor as a function of the frequency $\Omega$,  which denotes the frequency of  the external time-dependent electromagnetic  field resonantly driving the allowed transitions  between energy states of the system.   We have

\begin{eqnarray}
\sigma_{\alpha\beta}(\Omega)&=&\frac{N_fe^2}{2\Omega}
\int_{-\infty}^\infty \frac{d\omega}{2\pi} [ f(\omega-\mu)  -
f(\omega +\Omega -\mu) ]
\nonumber\\
&\times&  \int  \frac{d^3{\bf k}}{(2\pi)^3}
Tr\left[ \hat{v}_\alpha \hat{A}({\bf k},\omega+\Omega)   \hat{v}_\beta \hat{A}({\bf k},\omega )
\right] \  ,
\label{sigma}
\end{eqnarray}
where $N_f$ is a degeneracy factor for the spin, $e$ is the elementary charge, $f(\omega)$ is  the Fermi-Dirac distribution function, $\mu$ is the chemical potential and $\hbar \hat{v}_\alpha = \partial \hat{H}/\partial k_{\alpha}$  gives the velocity matrices $\hat{v}_\alpha$. Also, $\alpha,\beta$ represent spatial coordinates $x,y,z$.
For the longitudinal in-plane conductivity, we have $\hat{v}_x=-2k_x\hat{\sigma}_z$, $\hat{v}_y=-2k_y\hat{\sigma}_z$ and $\hat{v}_z=-2k_z\hat{\sigma}_z+\hat{\sigma}_x$. For temperature $T= 0$ K, the Fermi function acts as a step function so the integration limit changes to  $|\mu |- \Omega$ to $|\mu |$. In order to simplify the expressions, we chose  $E_k$ to represent  the magnitude of energy eigenvalue. Hence, we replaced $\epsilon_+ ({\bf  k})= E_k$ and $\epsilon_- ({\bf  k}) = -E_k $. Putting these results together, we conclude that  the longitudinal conductivity $\sigma_{xx}(\Omega)$, $\sigma_{yy}(\Omega)$ and $\sigma_{zz}(\Omega)$ are

\begin{eqnarray}
\sigma_{xx}(\Omega) &=& \frac{N_f e^2}{2 \hbar \Omega \pi^2} \int_{|\mu |- \Omega}^{|\mu |} d\omega  \int d^3 {\bf k} \ k^2_x
\nonumber\\
&\times & \left.
 \{ [\,1 - \frac{k^2_z}{E_k^2}]\,[\, \delta( \omega + \Omega - E_k ) \delta( \omega -E_k )+ \delta( \omega + \Omega + E_k) \delta( \omega +E_k )]\,\right.
\nonumber\\
&+&\left.
 \frac{k^2_z}{E_k^2}[\, \delta( \omega + \Omega - E_k ) \delta( \omega +E_k ) + \delta( \omega + \Omega + E_k ) \delta( \omega -E_k)]\, \}\right.
\label{numerics3}
\end{eqnarray}

\medskip
\par
\begin{eqnarray}
\sigma_{yy}(\Omega) &=& \frac{N_f e^2}{2 \hbar \Omega \pi^2} \int_{|\mu |- \Omega}^{|\mu |} d\omega  \int d^3 {\bf k} \ k^2_y
\nonumber\\
&\times & \left.
 \{ [\,1 - \frac{k^2_z}{E_k^2}]\,[\, \delta( \omega + \Omega - E_k ) \delta( \omega -E_k )+ \delta( \omega + \Omega + E_k) \delta( \omega +E_k )]\,\right.
\nonumber\\
&+&\left.
 \frac{k^2_z}{E_k^2}[\, \delta( \omega + \Omega - E_k ) \delta( \omega +E_k ) + \delta( \omega + \Omega + E_k ) \delta( \omega -E_k)]\, \}\right.
\label{numerics4}
\end{eqnarray}

\medskip
\par

\begin{eqnarray}
\sigma_{zz}(\Omega) &=& \frac{N_f e^2}{8\hbar \Omega \pi^2} \int_{|\mu |- \Omega}^{|\mu |} d\omega \, \int d^3 {\bf k} \  \{ [\,4k^2_z   \   (\, 1 - \frac{k_z^2}{E_k^2}-\frac{(\alpha-k^2)}{E_k^2} \, )\ \nonumber\\
 &+ & \left.\frac{k_z^2}{E_k^2}  ]\, \ [\, \delta( \omega + \Omega - E_k )  \ \delta( \omega -E_k )+ \delta( \omega + \Omega + E_k )  \ \delta( \omega +E_k )]\,\right.
 \nonumber\\
 &+ & \left.
  [\,4\frac{k_z^2}{E_k^2} \ [\,k_z^2 + (\alpha-k^2) ]\,+(1-\frac{k_z^2}{E_k^2})]\,  \ [\, \delta( \omega + \Omega - E_k ) \  \delta( \omega +E_k )\right.
\nonumber\\
 &+& \left.
 \delta( \omega + \Omega + E_k ) \  \delta( \omega -E_k)]\,\} \right.   \   . 
\label{numerics5}
\end{eqnarray}
 \medskip
\par
These terms give the intraband and interband contributions to the conductivity due to transitions within the conduction band and from transitions between the valence and conduction bands, respectively. The terms proportional to $\ [\, \delta( \omega + \Omega - E_k )  \ \delta( \omega -E_k )+ \delta( \omega + \Omega + E_k )  \ \delta( \omega +E_k )]\,$  vanish when $\Omega > 0$ and give only the intraband conductivity. Similarly, the terms proportional to $\ [\, \delta( \omega + \Omega - E_k )  \ \delta( \omega +E_k )+ \delta( \omega + \Omega + E_k )  \ \delta( \omega -E_k )]\,$ do not vanish for finite $\Omega$ and  give  interband conductivity. The significance of these contributions is determined by the level of doping, the frequency $\Omega$ and implicitly by $\alpha$ through the energy dispersion $\epsilon_{s}({\bf k})$.  The limit when  $\Omega \to 0$ corresponds to the Drude conductivity.\cite{Calvin} We analyzed these intraband (Drude) conductivity using a Lorentzian expression for the  $\delta$ function. 

\medskip
\par

\medskip
\par 
 Similarly, the transverse component of the conductivities are 
 
 \begin{equation}\sigma_{xy}  (\Omega) = \frac{N_fe^2}{2\Omega}
\int_{-\infty}^\infty \frac{d\omega}{2\pi} [ f(\omega-\mu)  -
f(\omega +\Omega -\mu) ] \ \int  \frac{d^3{\bf k}}{(2\pi)^3} \  (4 k_xk_y) \ {\bf\gamma}({\bf k},\omega ;\Omega) .
\label{numerics6}
\end{equation}


\begin{equation}
\sigma_{yz}  (\Omega) =   \frac{N_f e^2}{2\hbar \Omega} \int_{-\infty}^{\infty} \frac{d\omega}{2\pi}\;\left[ f(\omega - \mu)- f(\omega + \Omega -\mu)\right] \int  \frac{d^3 {\bf k}}{(2\pi)^3} \ \{-2k_y \ {\bf \kappa}({\bf k},\omega ;\Omega) + 4 k_y k_z  \ {\bf \gamma}({\bf k},\omega ;\Omega) \}
\label{numerics7}
\end{equation}
\medskip

 \begin{equation}
\sigma_{xz}  (\Omega)= \frac{N_f e^2}{2\hbar \Omega} \int_{-\infty}^{\infty} \frac{d\omega}{2\pi}\;\left[ f(\omega - \mu)- f(\omega + \Omega -\mu)\right] \int  \frac{d^3 {\bf k}}{(2\pi)^3} \ \{-2k_x \ {\bf \kappa}({\bf k},\omega ;\Omega) + 4 k_x k_z  \ {\bf \gamma}({\bf k},\omega ;\Omega)\ .
\label{numerics8}
\end{equation}
Clearly, the integrands in  Eq.\   (\ref{numerics6}),   \   (\ref{numerics7}) and \   (\ref{numerics8})  with values of ${\bf\gamma} $ and $ {\bf \kappa}$ from (\ref{Gamma}) and (\ref{Kappa}), all are an odd function of $k_x$ and $k_y$ thereby making $\sigma_{xy} (\Omega) $, $\sigma_{yz}  (\Omega)$ and $\sigma_{xz}  (\Omega)$  zero.

\medskip
\par

The expressions for the conductivity $\sigma_{xx}  (\Omega)$ and $\sigma_{yy}  (\Omega)$ are symmetric. Thus, we employ Eq.\   (\ref{numerics3}) and \   (\ref{numerics5}) in the following section  to carry out our numerical calculations for the longitudinal optical conductivity. These results exhibit the anisotropy of this NLSM system and demonstrate their dependence on the parameter $\alpha$ and the chemical potential $\mu$ as the frequency is varied.

\medskip
\par

\section{Results and Discussion For the Longitudinal Conductivity}
\label{sec3}

We now examine in detail with our numerical solutions the real part of the longitudinal optical conductivity of an anisotropic  NLSM  along the radial $x$, $y$ and axial  $z$ directions. For the calculations, we have used $N_f = 2$ corresponds to the spin degeneracy, the small momentum cutoff  with  $-1\leq k_{x,y,z} \leq 1$ to lie within the cubic region and a Lorentzian representation  for the Dirac delta function, i.e.,  \(\delta(x) \to \frac{\eta}{\pi} \frac{1}{\eta^2+x^2} \)  with a broadening of $\eta=0.05$.   This broadening takes into account the scattering due to non-magnetic impurities and lattice defects. The broadening is manifest in the optical conductivity as an effective transport scattering rate of  $\frac{1}{\tau}= 2\eta$. It is convenient to scale the conductivity by  $\sigma_0=e^2/4\hbar$ and we set $\hbar=1$  in the above Kubo formula. The finite temperature responses are also analyzed using the Fermi-Dirac distribution function at temperature $T$.

\begin{figure}[ht]
\centering
\includegraphics[width=0.5\linewidth]{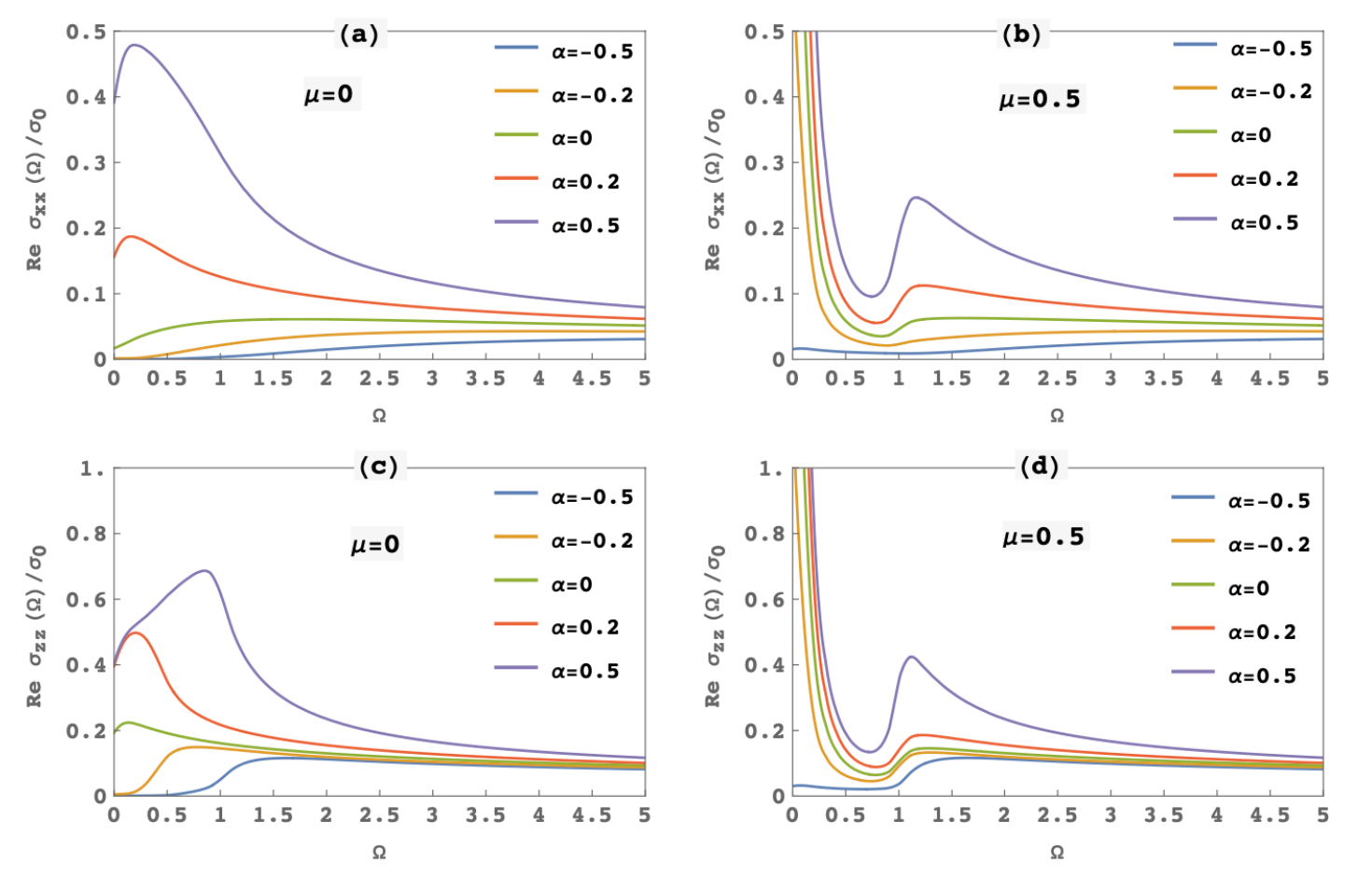}
\caption{ (Color online)   Real part of the longitudinal optical
conductivity at  T= 0 K  measured along the  $x$ and $z$ directions in the
top  to bottom panels, respectively, when (a) and (c)  for $\mu=0 $
and (b) and (d) for $\mu=0.5 $  for five chosen values of the tuning
 parameter  $\alpha $. As the value of $\alpha$ is
decreased, the conductivity becomes flatter and less dependent on
frequency. The given NLSM admits similar responses along $x$ and $y$
and slightly different response along the $z$ direction both
quantitatively and qualitatively. $\mu$, $\alpha$ and $\Omega$ are scaled with respect to energy unit $\hbar v_F k_c$.} 
\label{xx}
\end{figure}

\begin{figure}[ht]
\centering
\includegraphics[width=0.5\linewidth]{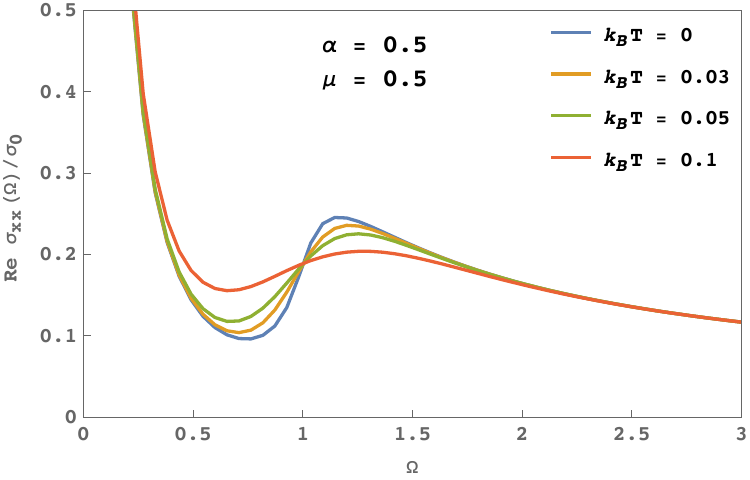}
\caption{ (Color online) The real part of the
optical conductivity along the $x-$axis is plotted as a function of
photon energy $\Omega$ for fixed chemical potential
$\mu = 0.5$ and adjustable  parameter $\alpha = 0.5$. The different curves corresponds to the various scaled temperature ($k_{B }T = 0, 0.03, 0.05, 0.1$) as indicated. Conductivity is normalized with $\sigma_0$ and $\mu$, $\alpha$, $\Omega$ and $k_{B}T$ are scaled in terms of energy unit $\hbar v_F k_c$.  When the temperature is increased from 0 K, the response slightly reduces and the curve becomes more flat at low frequency region.} 
\label{xxt}
\end{figure}

\medskip
\par
The transverse optical conductivity  vanishes because of rotational symmetry of the NLSM. The longitudinal conductivities along different directions are obtained accordingly with the band structure and density-of-states plots. The anisotropic optical responses of the NLSM are depicted in Fig.\ \ref{xx}  considering separately  two chosen values of $\mu$. The conductivities along $x$ and $y$ are similar. However, along the $z$ direction, the interband contribution for both pristine ($\mu = 0$) and the doped ($\mu=0.5$) NLSM has greater value. The delta peak near $\Omega$ equal to zero called intraband or Drude conductivity, is due to the transition of electrons within the conduction band which we have numerically calculated using the Lorentzian representation for the Dirac delta function in the expression. The non-zero value of optical conductivity even at zero frequency is due to the topologically protected gapless states on the nodal line. This feature of the optical frequency makes the topological NLSM different from narrow band gap semiconductor, where the optical conductivity is typically zero at zero frequency. These properties indicate possible applications involving microelectronics technologies. These could  include integrated circuits, ultrafast modulators and high performance transistors, as well as the development of self-detection of current-surge-induced overheating in electronic devices.

\medskip
\par
In the case of finite  $\mu$, we have zero conductivity for frequency less than $2 \mu$ and a finite jump for frequency greater than $2 \mu$ which reveals that the conductivity is  due to interband transitions and that transitions obey the Pauli exclusion principle. For photon energy greater than $2\mu$, the conductivity reaches a maximum value and then is decreased to a minimum saturated value. The peak of the curve depends significantly on the tunable parameter $\alpha$ as well as the chemical potential $\mu$. The conductivity becomes flat and the strength is reduced  when $\alpha$ continuously goes from finite positive value to negative value. Specifically, the  intensity of optical conductivity is reduced when a sizeable gap is opened by tuning the value of $\alpha$. On the other hand, the spectral weight of the conductivity tensor seems to be inversely dependent on the chemical potential. As we increase the chemical potential from zero, the intraband conductivity near zero frequency sharply rises but the strength of the  optical absorption is reduced . Therefore for the further increase in the value of $\mu$, the absorption peak move to the higher frequency region and the height also reduces. It means the electrons deep inside the valance band excite to the conduction band which require more photons with enough energy. Carbotte and Mukherjee. \cite{carbotte, mukherje} have reported the flat conductivity in NLSM with notable  finite height. However, our results are similar  to that for  ZrSiS. \cite{kohino}   In fact, Habe and Koshino demonstrated that the optical conductivity of a NLSM such as ZrSiS    takes the peak value and then sharply falls to a flat plateau region and again rises up in the high frequency region. Throughout these calculations, we have considered a toy-model Hamiltonian with only two subbands and small momentum cutoff.  Consequently, these results are applicable for small values of  chemical potential and for low frequency. For large values of chemical potential and for higher frequency range, the contribution from other subbands to the conductivity may be  significant and the conductivity may again increase. The effect due to finite temperature manifests itself through similar qualitative behavior. Figure\  \ref{xxt} demonstrates that when the temperature is increased from 0 K, the magnitude and the slope of the  interband optical conductivity curve is slightly reduced. This is because the thermally excited electrons and holes are perturbed by  the optical transitions from the valance to the conduction band in the low-frequency regime. When the temperature is further increased from room temperature, the curve becomes more flat. This means that the effect due to temperature is not as significant as it is  in the low-temperature regime. We generalized this finite temperature response for the $4 \times 4$ matrix model Hamiltonian  which we describe below.

\section{Density-of-states}
\label{sec4}

\medskip

\begin{figure}[ht]
\centering
\includegraphics[width=0.5\linewidth]{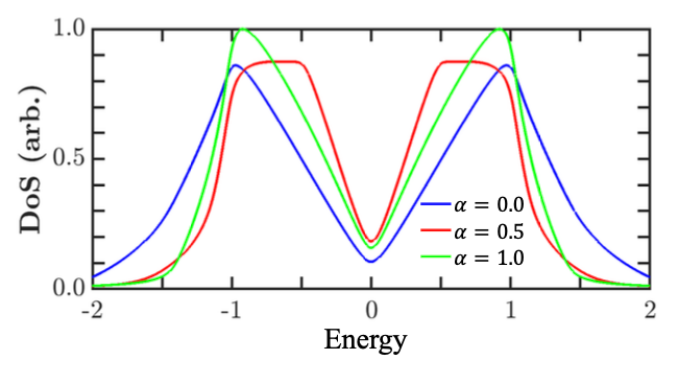}
\caption{(Color online)   Density-of-states for chosen values of
$\alpha$ . Energy and $\alpha$ are scaled with respect to  $\hbar v_F k_c$.} 
\label{dos}
\end{figure}

For the density-of-states, we have

\begin{equation}
N(\omega)= N_f \int  \frac{d^3{\bf k}}{(2\pi)^3} \sum_{s=\pm} \delta (\omega -\epsilon_s({\bf k}))
\end{equation}
for which we again employ a Lorentzian representation for the delta  function. The effects due to the gaps between the energy subbands that are generated by the positive  parameter   $\alpha $ are clearly reflected in the density-of-states.  The presented density-of-states in Fig.\ \ref{dos} are  for three values of $\alpha=0,\ 0.5$ and $1.0$

\section{Another example of a nodal-line semimetal}
\label{sec5}

 An effective Hamiltonian for electrons in
nodal-line semimetals with $Z_{2}$ indices $\zeta_{1} = 1$ and
$\zeta_{2} = 1$ is given by~\cite{ChiJou}

\begin{eqnarray}
\label{Ham2}
\hat{H} (\mathbf{k}) = k_{x}s_{x}\otimes I + k_{y}s_{y}\otimes
\tau_{y} + k_{z}s_{z} \otimes I + \alpha \tau_{x}\otimes s_{x} ,
\end{eqnarray}
where $\otimes$ is the Kronecker product of two matrices. 
For the model Hamiltonian~(\ref{Ham2}), by varying the value of $\alpha$ from positive to negative, the nodal line on the $k_x-k_y$ plane of radius $|\alpha |$, first shrinks to a point, then evolves into a ``line" again. This means that one cannot open a gap between the valance and conduction band  just by varying the parameter $\alpha$. The shape and extent of the nodal line change symmetrically for positive and negative $\alpha$. The detailed analysis of topology of NLSMs is provided in
Ref.~[\onlinecite{ChiJou}]. The Hamiltonian $\hat{H} (\mathbf{k})$
in Eq.~(\ref{Ham2}) reads in matrix form as

\begin{eqnarray}
 \label{Ham4mat}
\hat{H} (\mathbf{k}) = \left(
\begin{array}{cccc}
k_{z} & 0 & k_{x} & \left(\alpha - k_{y}\right) \\
0 & k_{z} & \left(\alpha + k_{y}\right) & k_{x}  \\
k_{x} & \left(\alpha + k_{y}\right) & - k_{z} & 0  \\
\left(\alpha - k_{y}\right)& k_{x} & 0 & - k_{z}
\end{array}%
\right)  .
\end{eqnarray}
The eigenvalues $\epsilon_{s,s^\prime} ({\bf k})$ of $\hat{H} (\mathbf{k})$ from
Eq.~(\ref{Ham4mat}) are

\begin{eqnarray}
\epsilon_{s,s^\prime} ({\bf k})= s \sqrt{k_{z}^{2} + \left(\sqrt{k_{x}^{2} +
k_{y}^{2}} \ + s^\prime \alpha  \right)^{2}}  \    .
\label{E1234}
\end{eqnarray}

\begin{figure}[ht]
\centering
\includegraphics[width=0.5\linewidth]{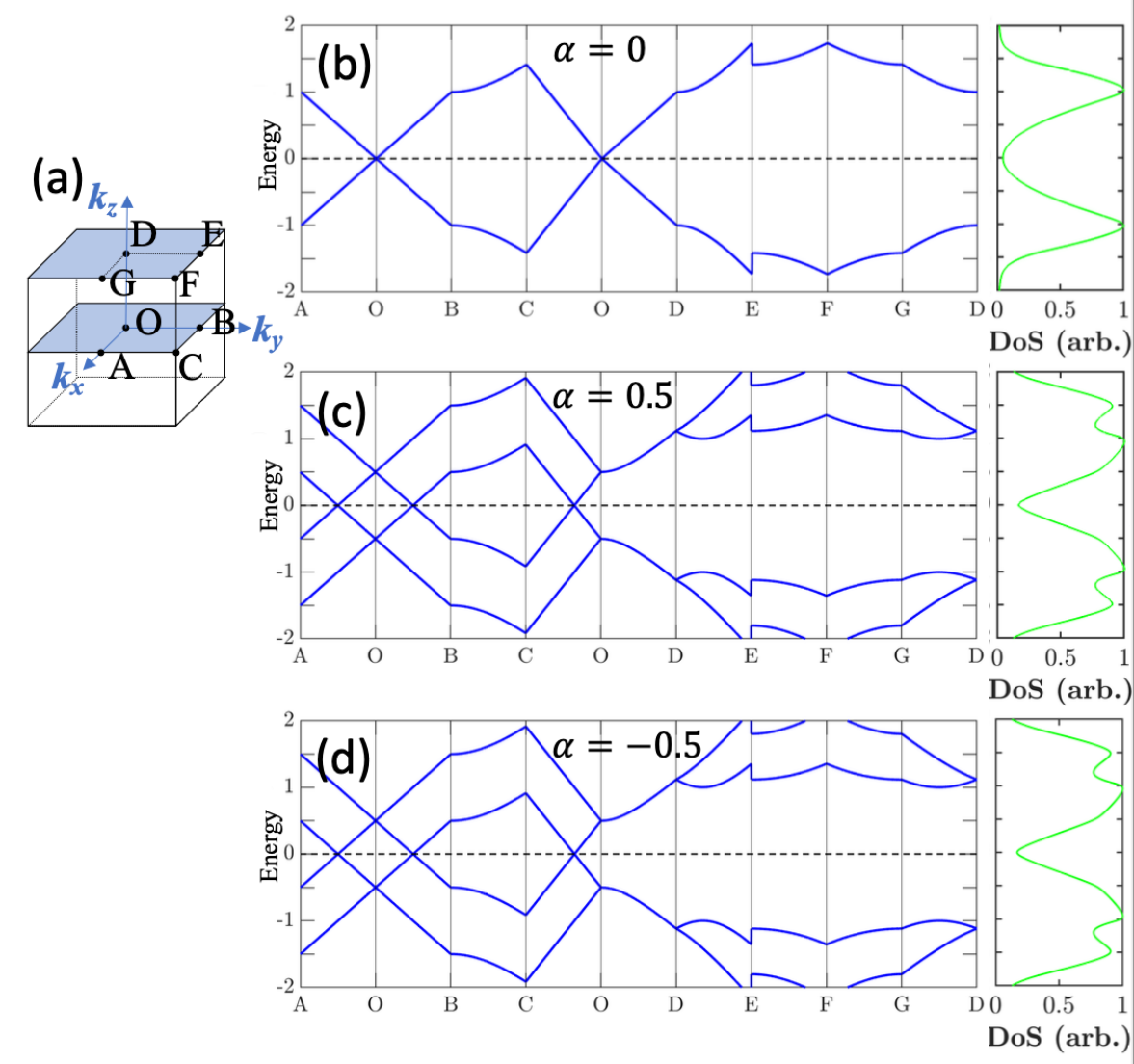}
\caption{ (Color online) (a)   Low-energy subbands of bulk NLSM described by the Hamiltonian in Eq.\ (\ref{Ham2}). The results are within  a cubic region  with indicated symmetry points. Bulk bands along  lines joining the given symmetry points (left) and corresponding density-of-states (right) for (b)   $\alpha=0$, (c) $\alpha=0.5 $ and (d) $\alpha=-0.5 $. The bands and density-of-states are symmetric for $\alpha = 0.5$ and $\alpha=-0.5 $.}
\label{B3}
\end{figure}

\begin{figure}[ht]
\centering
\includegraphics[width=0.5\linewidth]{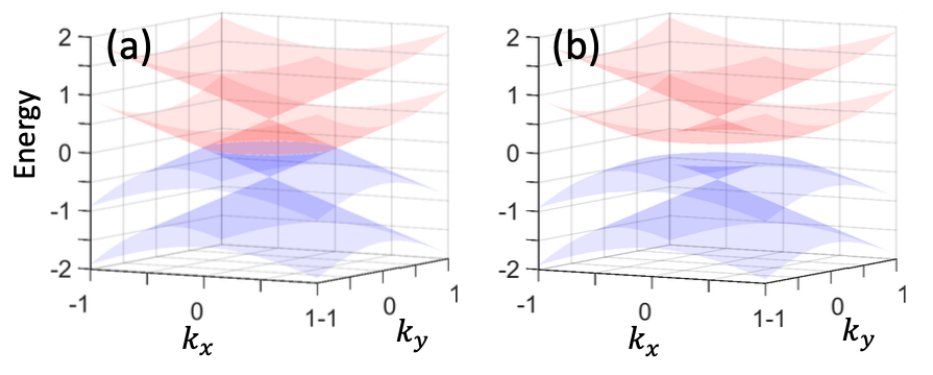}
\caption{ (Color online)  The 3D contour bands at low energy on the selected ($k_x,k_y$) planes for (d)  $k_z=0$ and (e)  $k_z= 0.2$ at $\alpha=0.5$. The pink curve corresponds to the conduction band ($s=+$) and the blue curve corresponds to the valence band ($s= -$ )  }
\label{B4}
\end{figure}

\medskip
\par

\medskip
\par
Also, we have determined the eigenvectors for this case but the expressions are long and unwieldy and not needed for calculating the optical conductivity. We will need these for determining the plasmon dispersion. For calculating $\sigma_{\alpha\beta}(\Omega)$, we employ

\begin{equation}
 \label{xvz}
\hat{v}_x=    \left(
\begin{array}{cccc}
0 & 0 & 1 & 0 \\
0 & 0 & 0 & 1 \\
1 & 0 &  0 & 0  \\
0 & 1 & 0 & 0
\end{array}%
\right)  \ , \  \  \    \
\hat{v}_y=    \left(
\begin{array}{cccc}
0 & 0 & 0 & -1 \\
0 & 0  & 1 & 0 \\
0 & 1 & 0 & 0  \\
-1 & 0 & 0 &  0
\end{array}%
\right)  \ , \  \  \    \
\hat{v}_z=    \left(
\begin{array}{cccc}
1 & 0 & 0 & 0 \\
0 & 1 & 0 & 0 \\
0 & 0 & - 1 & 0  \\
0 & 0 & 0 & - 1
\end{array}
\right)    \
\end{equation}

\begin{eqnarray}
\epsilon_1 &=&\epsilon_{++} = \sqrt{k^2_z + (\sqrt{k^2_x + k^2_y}+ \alpha)^2}\nonumber \\
\epsilon_2 &=& \epsilon_{+-} = \sqrt{k^2_z + (\sqrt{k^2_x + k^2_y}- \alpha)^2}\nonumber \\
\epsilon_{-+} &=& - \epsilon_1 \ and  \ \epsilon_{--} =  -\epsilon_2 \nonumber \\
\label{eigenvalue2}
\end{eqnarray}

\medskip
\par
In this case, the Green's function and the spectral function have the matrix elements in the form 
\begin{eqnarray}
 \hat{{\cal G}}_{11}(\omega)&=&\frac{1}{{\cal  D}_2(\omega)}\left\{
-\alpha^2k_z-\omega\alpha^2 -2\omega k_yk_z -2\omega\alpha k_y -k_x^2k_z -\omega k_x^2 -k_y^2k_z -\omega k_y^2 -k_z^3-\omega k_z^2 +\omega^2 k_z +\omega^3\right\}
\label{App2}
\end{eqnarray} 
And \begin{eqnarray}
A_{11}(\omega) &=& \frac{\pi }{4\alpha} \frac{\omega+k_z}{\sqrt{k^2_x + k^2_y}}\ (\omega^2 - k^2_z - k^2_x - k^2_y - \alpha^2 -2\alpha k_y)\left\{ \frac{1}{\epsilon_1}( \delta(\omega-\epsilon_1) -
 \delta(\omega+\epsilon_1))- \frac{1}{\epsilon_2}( \delta(\omega-\epsilon_2) -
 \delta(\omega+\epsilon_2))
\right\}
\nonumber\\
\label{spectralfn2}
\end{eqnarray}
The other elements have the similar expressions. The function ${\cal  D}_2(\omega)=(\omega-\epsilon_{+,+})(\omega -\epsilon_{+,-})(\omega -\epsilon_{-,+})(\omega -\epsilon_{-,-})$ where the eigenvalues are given in Eq.\ (\ref{E1234}). Using the Kubo formula Eq.\ (\ref{sigma}), the real part of longitudinal optical conductivity is obtained as

\begin{equation}
\sigma_{xx}(\Omega) = \frac{N_f e^2}{2  \hbar  \Omega  (2\pi)^4} \int_{|\mu |- \Omega}^{|\mu |} d\omega  \int  d^3 {\bf k}  \ {\bf\Delta}({\bf k},\omega ;\Omega)   \
\label{sigmax2}
\end{equation}\
\begin{equation}
\sigma_{yy}(\Omega) = \frac{N_f e^2}{2  \hbar  \Omega  (2\pi)^4} \int_{|\mu |- \Omega}^{|\mu |} d\omega  \int  d^3 {\bf k}  \ {\bf\Sigma}({\bf k},\omega ;\Omega)   \
\label{sigmay2}
\end{equation}\
\begin{equation}
\sigma_{zz}(\Omega) = \frac{N_f e^2}{2  \hbar  \Omega  (2\pi)^4} \int_{|\mu |- \Omega}^{|\mu |} d\omega  \int  d^3 {\bf k}  \ {\bf\Lambda}({\bf k},\omega ;\Omega)   \
\label{sigmaz2}
\end{equation}

where, the expressions for ${\bf\Delta}({\bf k},\omega ;\Omega)$, ${\bf\Sigma}({\bf k},\omega ;\Omega)$ and ${\bf\Lambda}({\bf k},\omega ;\Omega)$ are given in the Appendix C.

\begin{figure}[ht]
\centering
\includegraphics[width=0.5\linewidth]{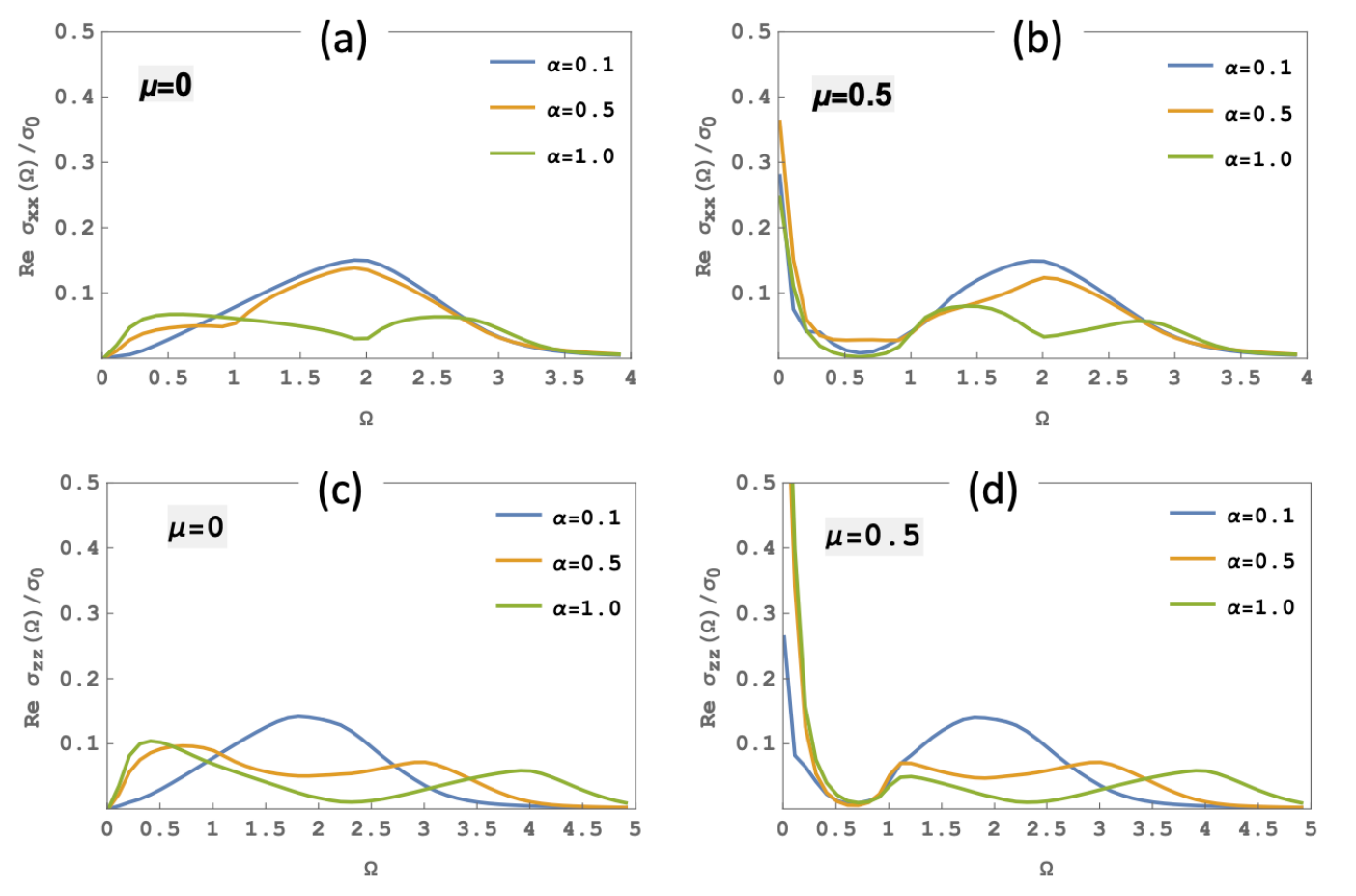}
\caption{ (Color online)   Real part of  the longitudinal optical conductivity of nodal-line semimetal described by the model Hamiltonian  in Eq.\  (\ref{Ham2}) at T=0 K,  (a) and (b) along the $x$ direction and (c) and (d)  along the $z$ direction  when (a) and (c) for $\mu=0 $  and (b) and (d)  for $\mu=0.5 $  with chosen values of the tuning parameter $\alpha $. $\mu$, $\alpha$ and $\Omega$  are scaled with  respect to energy unit $\hbar v_F k_c$.}
\label{xx2}
\end{figure}

\medskip
\par
Needless to say, a single model Hamiltonian  cannot explain all types of semimetals. Different model Hamiltonians can be used to understand various materials of NLSM groups. This is exemplified in the work of these two references \  [\onlinecite{1PRBRef2}] and \  [\onlinecite{2PRBRef2}] . It depends on the shape, size and nature of the nodal line. Since the radius of the nodal ring can vary, depending on the strength of the tuning parameter, we have investigated the optical responses of such NLSM whose nodal  ring does not shrink to a point and is gapped by tuning the parameter $\alpha$. We have characterized the optical conductivity considering the effective Hamiltonian in Eq. (\ref{Ham2}) which has four eigenvalues corresponds to four subbands in the low-energy cubic region. Two valence bands and two conduction bands are clearly outlined in the band structure plot in Fig.\  \ref{B3}. When the value of $\alpha$ is set to zero, only two subbands are present within this cubic region and the subbands are symmetric along three axes and are linearly dispersed around the $O$ point as that for Dirac or Weyl semimetals. However, for finite value of $\alpha$, two subbands are present within the this cubic region and they are not symmetrically dispersive as we go from $O$ to $A$ or $B$ and $O$ to $D$. We are more focused on the results for finite $\alpha$.  The conduction and  valence bands are gapped  in parallel planes for  $k_z$ not equal to zero as shown in  Fig.\  \ref{B4}. The density-of-states also has two peaks and has finite value for larger range of energy than  the previous model.

\medskip
\par
The optical conductivities presented in Fig.\  \ref{xx2} are in accordance with the band structure and density-of-states for the Hamiltonian in Eq. (\ref{Ham2}). Various types of intraband and interband transition channels are allowed.  During the process of our calculations, we have set $N_f=2$ to include the spin degeneracy of the bands.  For  charge neutral pristine NLSM, zero conductivity at $\Omega \rightarrow 0$ reveals that there are no allowed transitions within the subbands. Therefore, the total conductivity for $\mu=0$ is the contribution from interband transitions. As the Fermi level rises up to a positive value, intraband conductivity contributions are achieved and delta peaks appear around $\Omega$ tending to zero. Unlike the previous model Hamiltonian, the conductivity is inversely proportional to $\alpha$. As the value of  $\alpha$ is increased from $0.1$, the height of the single peak is reduced and is split into two reduced peaks for $\alpha$ around $1.0$. Eventually, all curves saturates to a minimum value.  The nature of the curves is slightly different in the upper ($\sigma_{xx}$) and lower ($\sigma_{zz}$) plots. For larger values of $\alpha$, the curves are flatter along $z$ than that along $x$ which reveals the anisotropy of the conductivity.

\medskip
\par
 \section{Thermal Properties}
 \label{sec6}

We employ the Wilson-Sondheim formula  \cite{Ref13,Ref14}  for the grand thermodynamic potential of a Fermi-Dirac gas of noninteracting electrons.  We have per unit volume

\begin{eqnarray}
\Omega (T,\mu ) &=&\  -k_BT \frac{1}{V}\sum_{{\bf k},s}
\ln\left[1+e^{-\beta(E_{{\bf k},s}-\mu)}  \right]
\end{eqnarray}
where $V$ is a normalization volume and $E_{{\bf k},s}$ is  energy eigenvalue given in Eq. (2). The calculated grand potential can be further used to calculate several thermodynamic quantities for various temperatures and densities.    It is worthy of note that  $\Omega(T,\mu)$ may also be expressed in terms of the inverse Laplace transform of the partition function \cite{Ref13,Ref14}   \( Z(s)=\sum_{\{\omega\}} e^{-s\omega} \), where the sum is carried out over the eigenstates, with an appropriate cut-off for the wave vector.  The specific heat capacity  is

\begin{eqnarray}
C_V   &=&  T\frac{\partial S}{\partial T}
= - \beta \frac{\partial S}{\partial \beta}    
\nonumber\\
&=&    \frac{k_B \beta^2}{2}   \frac{1}{V}\sum_{{\bf k},s}  \frac{ (E_{{\bf k},s}-\mu)^2}{\cosh^2(\beta(E_{{\bf k},s}-\mu))/2}
 \end{eqnarray} 
which follows from the entropy

\begin{eqnarray}
S   &=&  k_B\beta^2  \frac{\partial\Omega}{\partial\beta}
\nonumber\\
&=&    k_B   \frac{1}{V}\sum_{{\bf k},s}   \ln\left[1+e^{-\beta(E_{{\bf k},s}-\mu)}  \right]
+k_B \beta  \frac{1}{V}\sum_{{\bf k},s}   (E_{{\bf k},s}-\mu) \left\{ 1-\tanh\left( \frac{\beta (E_{{\bf k},s}-\mu)}{2} \right)
\right\}
\end{eqnarray} 

\medskip
\par
\begin{figure}[ht]
\centering
\includegraphics[width=0.5\linewidth]{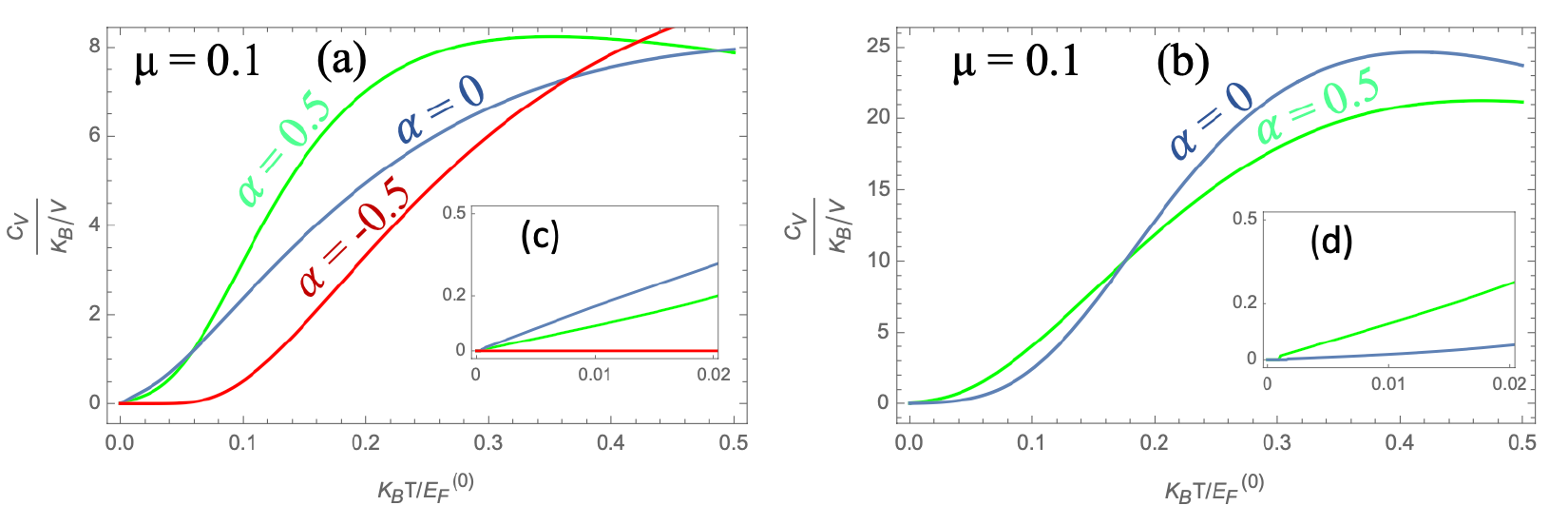}
\caption{ (Color online)  Temperature variation of specific heat capacity at constant volume (a) for the model Hamiltonian  in Eq.\  (1) and (b) for the model Hamiltonian  in Eq.\  (14) for $\mu=0.1 $. (c) and (d) are the  low-temperature portions  of (a) and (b) respectively. The different curves are for  different values of the tuning parameter $\alpha $ as indicated. $C_{V}$ is normalized to constant value $K_{B}/V$ and $K_{B}T$ is normalized to  energy $E_{F}^{(0)} = \hbar v_{f}k_{c}$.  }
\label{capacity}
\end{figure}
\medskip
\par
Numerical results for the specific heat capacity are presented in Fig.\ \ref{capacity}.  Plots (a) and (b) are for the two model Hamiltonians considered in Eqs.\  (1) and    (14). For Fig. 9 (b), the energy is given by Eq.\ (16) and the summation is taken over all subbands $s$ and $s^{\prime}$. We have found that the heat capacity is low at very low temperatures. As the temperature is increased, it rises linearly with temperature and attains a saturated value following the Debye/Einstein law. However, the saturation temperature and low-temperature response are strongly dependent on the tuning parameter $\alpha$. The heat capacity variation with temperature around room temperature is  shown in the zoom-in part of each plot. It shows that, the  heat capacity with tuning parameter $\alpha$ in the low temperature region is different for the two model Hamiltonians considered. For instance, in Fig.\ \ref{capacity}(c), the curve for $\alpha = 0$ has higher slope than that for $\alpha = 0.5$ and for $\alpha = -0.5$ it is zero. On the other hand, in Fig.\ \ref{capacity}(d), the curve for $\alpha = 0$ has lower slope than that for $\alpha = \pm0.5$. It signifies that one can vary the specific heat capacity by tuning the gap in the NLSM.


\section{Summary and  Conclusions}
\label{sec7}

In conclusion,  we  have theoretically investigated the linear responses of nodal-line semimetal using two simple models in conjunction with the Kubo formula for the optical  conductivity as well as their heat capacity when in thermal equilibrium.  We have derived closed-form expressions for the longitudinal and transverse components of the dynamical optical conductivity for both of these simple models  whose energy subband structures are anisotropic in ${\bf k}$ space.   It has been found that NLSM  presents different thermal behavior depending on the gap in the band structure.  We may also employ the grand potential to calculate other  thermodynamic properties such as compressibility, expansion coefficient and sound velocity. These model Hamiltonians are simple yet their energy subband structures bear similarities to calculated results using the tight-binding model near the Fermi level of a nodal-line semimetal under strain. This effect due to strain is introduced through a parameter $\alpha$ in the model Hamiltonian, thereby making these simple models useful for predictions for the NLSM using this approach. This is bolstered not only by  similarities in the band structure but the overall qualitative behavior of the conductivity when compared with a material such as ZrSiS.  We have presented  results for the optical conductivity as a function of  frequency  at both zero and finite temperature. We analyzed the results considering the effect of varying  the parameter $\alpha$ for two different models of nodal-line semimetal and chemical potential.  We have plotted the subband structure along lines joining  given symmetry points in a low-energy cubic region in a chosen $(k_{x}, k_{y})$ plane.  We emphasize that the continuum model Hamiltonian  yields energy  bands near the $O$ point in the $(k_{x}, k_{y})$  plane but these results could be quite different in parallel planes.  As we go from some positive value of $\alpha$ to zero, the crossing of the valence and conduction bands changes from nodal ring to 3D point node. However, for negative value of $\alpha$, the dispersion is different for the two models. For the  $2 \times 2$ matrix model Hamiltonian, the valence and conduction bands separate and a gap opens near the $O$ point while  for the $4 \times 4$ matrix model Hamiltonian they again cross, making a nodal ring. This means that one cannot tune one Hamiltonian to the other without closing the gap. Therefore, these two model Hamiltonians are not topologically equivalent and hence the drastic difference in the optical conductivity of these two models signifies that the optical responses of such NLSM are sensitive to topology. Additionally, we do not expect  similar features in a trivial semiconductor with tunable band gap  because the band structure is different from the ones obtained for the two model Hamiltonians for NLSM, presented in this paper. The specific heat capacity is also different for the two models.  Furthermore, the results we obtained for the dynamical optical conductivity are in qualitative agreement with those reported using realistic band structures. In the Table 1, we have listed the references in which the optical response of different NLSM materials has  been studied experimentally and theoretically.

\medskip
\par

Table 1: The following references studied the optical conductivity
of NLSM materials experimentally and theoretically.

\begin{table}[h]
  \begin{tabular}{|c|c|}
      \hline
      \textbf{  Reference} & \textbf{Case study}     \\
      \hline
      [6] &  Cd$_{3}$As$_{2}$, TaAs, Eu$_{2}$In$_{2}$O$_{7}$, ZnTe$_{5}$  \\
      \hline
     [7]  & nodal loop semimetals  \\
    \hline
     [8] &  body-centered orthorhombic C$_{16}$, CaP$_{3}$, Ca$_{3}$P$_{2}$   \\
    \hline
    [9] &  ZrSiS, ZrSiSe, ZrSiTe,  HfSiS  \\
    \hline
    [11] & tilted NLSM  \\
    \hline
    [12] &  ZrSiS  \\
      \hline
    [13] &  YbMnSb$_{2}$ \\
    \hline
    [14] & ZrSiS, ZrSiSe , ZrSiTe, ZrGeS, ZrGeTe     \\
    \hline
    [15] &   ZrGeS, ZrGeSe   \\
    \hline
    [16] &  ZrSiS              \\
    \hline
  \end{tabular}
\end{table}

\medskip
\par
The transverse conductivity vanishes due to rotational symmetry along the  nodal ring of the semimetal.  Regarding the longitudinal component, for charge neutral NLSM, the interband conductivity is significant near zero frequency. However, for doped NLSM with chemical potential $\mu$,  the interband conductivity rises to the peak value at $\Omega=2\mu$ and is then decreased. For frequency less than $2\mu$, Pauli blocking makes electrons unable to make  transitions  in the valence and conduction bands. At higher photon energies, the transition saturates at its minimum value. This can be explained as a consequence of the  low-energy cut-off and the resulting limits on the wave vector ${\bf k} = (k_x, k_y, k_z)$ scaled by $k_c$ which is restricted to $| k_i | \leq \ 1$ ($i=x,y,z$). The Hamiltonian $\hat{H}$, $\Omega$, $\alpha$ and $\mu$ are all scaled in terms of the energy unit $\hbar v_F k_c$. Therefore, the conductivity of NLSM depends on  $v_F $ and $k_c$. The height and width of the peak depend principally on the tuning parameter ($\alpha$). For $\alpha=0$, the curve is almost flat at very small transition. At $\alpha =0$, the bands touch at a point but the quadratic dispersion near $O$ is not linear like Dirac or Weyl. Therefore, the conductivity is not flat like that of 2D and 3D Dirac materials as explained in Refs.\   [\onlinecite{carbotte, barati, mukherje}].   As we increase the value of $\alpha$ from $0$ to larger values, the peak becomes sharper and sharper. In any event, they all converge to a minimum saturated value in the higher frequency region. This is also in agreement with density-of-states plots. The density-of-states has a peak at a certain energy and is zero over a range which could be adjusted by manipulating the value of $\alpha$. The anisotropic dispersion relation along radial ($k_x, k_y$) and axial ($k_z$)  directions results in their anisotropic conductivity along these directions. The conductivity along $x$ and $y$ are identical yet the conductivity along $z$ deviates slightly both quantitatively and qualitatively. The anisotropic conductivities in Shahin's \cite{barati} paper are explained as a consequence of an anisotropic Fermi surface for which they have considered a toroidal shape.

\medskip
\par
The nodal lines in these model Hamiltonians are protected by three symmetries such as inversion symmetry, time-reversal symmetry and spin rotation $SU(2)$ symmetry. Since the SOC has not been considered, these symmetries are preserved under the application of strain. When SOC is introduced, these symmetries can be broken leading to the splitting of the line node into point nodes. This splitting of the line node into point node may affect the band structure and the density of states. Moreover, it can enhance light-matter interactions and give rise to new optical transitions. Therefore, we can expect the different features like changes in strength, changes in anisotropy as well as the appearance or disappearance of peaks in the spectrum of optical conductivity. A detailed analysis of optical conductivity in the presence of SOC is interesting in its own right.

\medskip
\par
 Although the model Hamiltonians used undoubtedly have limitations, we believe that our work  serves to complement the existing literature with our reported results for optical conductivity as well as thermal heat capacity  and their dependence on the gap, temperature and doping. This will serve to advance our knowledge beyond the already known results. Additionally, there are several applications of NLSMs. Topological nodal-line semimetals with many exotic physical properties are potential candidates for next-generation device applications. For example, the high photosensitivity is of importance for the construction of ultrafast photodetectors. The spin-momentum locking of the surface states could be used for low-consumption spintronic devices and magnetic memory devices. The nodal lines of the semimetal can be used to create protected qubits for the development of topological quantum computing. In addition, many topological nodal-line semimetals have large thermoelectric responses, which can be utilized for high-efficiency energy converters or thermal detectors.

\medskip
\par

\medskip
\pagebreak

\section*{Acknowledgement(s)}
G.G. would like to acknowledge the support from the Air Force Research Laboratory (AFRL)
through Grant No. FA9453-21-1-0046.   We would like to thank Dr. Po-Hsin Shih and Dr. Thi-Nga Do  for helpful discussions and guidance on the numerical calculations  and interpretation of the results.

\medskip
\par

\appendix

\section{Spectral function}

We have from Eq.\   (\ref{grnfn})  and setting $i=j=1$

    $$ G_{11}(z) = \int_{-\infty}^{\infty} \frac{d\omega^{'} A_{11}(\omega^{'})}{2\pi (z-\omega^{'})}\\$$
	
   Resolving the left-hand side of this equation into    partial fractions, we obtain
			 $$\frac{1}{(\epsilon_{+}-\epsilon_{-})} \left[ \frac{[\epsilon_{+}+ (\alpha - k^{2})]}{(z-\epsilon_{+})} - \frac{[\epsilon_{-}+ (\alpha - k^{2})]}{(z-\epsilon_{-})}) \right]=\int_{-\infty}^{\infty} \frac{d\omega^{'} A_{11}(\omega^{'})}{2\pi (z-\omega^{'})} $$ 
Now, let us rewrite the left-hand side of this equation in integral form  which is possible when $\delta$ functions are introduced. The result is 
$$\frac{1}{(\epsilon_{+}-\epsilon_{-})}\int_{-\infty}^\infty
\frac{d\omega^\prime}{z-\omega^\prime}
\left[ [\epsilon_{+}+ (\alpha - k^{2})] \ \delta(\omega^\prime-\epsilon_{+}) - [\epsilon_{-}+ (\alpha - k^{2})]  \ \delta(\omega^\prime -\epsilon_{-}) \right] =\int_{-\infty}^{\infty} \frac{d\omega^{\prime} A_{11}(\omega^{\prime })} {2\pi (z-\omega^{\prime }) }    \  .$$
From this latter equation,  we clearly identify the spectral weight $A_{11}(z)$as

 $$A_{11}(z)= \frac{2\pi}{(\epsilon_{+}-\epsilon_{-})} \ \left[ [\epsilon_{+}+ (\alpha - k^{2})] \ \delta(z-\epsilon_{+}) - [\epsilon_{-}+ (\alpha - k^{2})]  \ \delta(z-\epsilon_{-}) \right]\   . $$

\medskip
\par

\section{Conductivity I}

The following are the expressions for ${\bf\gamma}({\bf k},\omega ;\Omega)$ and ${\bf\kappa} ({\bf k},\omega; \Omega)$ in  Eq.\   (\ref{numerics6}),   \   (\ref{numerics7}) and \   (\ref{numerics8}) .





\begin{eqnarray}
{\bf\gamma}({\bf k},\omega ;\Omega)    &  \equiv&   
 4 \pi^2\{ [\,1 - \frac{k^2_z}{E_k^2}]\,[\, \delta( \omega + \Omega - E_k ) \  \delta( \omega -E_k )+ \delta( \omega + \Omega + E_k) \  \delta( \omega +E_k )]\,\nonumber\\
&+&\left.
 \frac{k^2_z}{E_k^2}[\, \delta( \omega + \Omega - E_k) \  \delta( \omega +E_k ) + \delta( \omega + \Omega + E_k) \  \delta( \omega -E_k)]\, \}\right. \ ,
\label{Gamma}
\end{eqnarray}

\begin{eqnarray}
{\bf\kappa} ({\bf k},\omega; \Omega)  &\equiv&   
 \frac{4\pi^2 k_z (\alpha-k^2)}{E_k^2} \ \{ \delta(\omega+\Omega - E_k) \ \delta(\omega-E_k)- \delta(\omega+\Omega + E_k) \ \delta(\omega-E_k)
\nonumber\\
&-&\left.
 \delta(\omega+\Omega - E_k) \  \delta(\omega+E_k)+ \delta(\omega+\Omega + E_k) \ \delta(\omega+E_k)\}\right.
\label{Kappa}
\end{eqnarray}


\section{Conductivity II}
 The expressions for ${\bf\Delta}({\bf k},\omega ;\Omega)$, ${\bf\Sigma}({\bf k},\omega ;\Omega)$ and ${\bf\Lambda}({\bf k},\omega ;\Omega)$ in Eq. (\ref{sigmax2}), \ (\ref{sigmay2}) and (\ref{sigmaz2}) are given by,

\begin{eqnarray}
{\bf\Delta}({\bf k},\omega ;\Omega)  & = &   A_{11}(\omega+\Omega)  \ A_{33}(\omega)+ A_{33}(\omega+\Omega)  \ A_{11}(\omega) + A_{22}(\omega+\Omega)  \ A_{44}(\omega) + A_{44}(\omega+\Omega)  \ A_{22}(\omega) \nonumber\\
  &+& \left. 2 A_{12}(\omega+\Omega)  \ A_{34}(\omega)+ 2 A_{34}(\omega+\Omega)  \ A_{12}(\omega) + 2 A_{14}(\omega+\Omega)  \ A_{23}(\omega) + 2 A_{23}(\omega+\Omega)  \ A_{14}(\omega)\right. \nonumber\\
  & + & \left. 4 A_{13}(\omega+\Omega) \ A_{13}(\omega)\right.
\nonumber\\
\label{delta}
\end{eqnarray}
\begin{eqnarray}
{\bf\Sigma}({\bf k},\omega ;\Omega)  & = &  -2 A_{12}(\omega+\Omega)  \ A_{34}(\omega)-2 A_{34}(\omega+\Omega)  \ A_{12}(\omega) + A_{22}(\omega+\Omega)  \ A_{33}(\omega) + A_{33}(\omega+\Omega)  \ A_{22}(\omega) \nonumber\\
  &+& \left. 2 A_{23}(\omega+\Omega)  \ A_{23}(\omega)+ 2 A_{14}(\omega+\Omega)  \ A_{14}(\omega) +  A_{44}(\omega+\Omega)  \ A_{11}(\omega) +  A_{11}(\omega+\Omega)  \ A_{44}(\omega)\right. \nonumber\\
  & - & \left. 4 A_{13}(\omega+\Omega) \ A_{13}(\omega)\right.
\nonumber\\
\label{sum}
\end{eqnarray}
\begin{eqnarray}
{\bf\Lambda}({\bf k},\omega ;\Omega)  & = &   A_{11}(\omega+\Omega)  \ A_{11}(\omega)+ A_{22}(\omega+\Omega)  \ A_{22}(\omega) + A_{33}(\omega+\Omega)  \ A_{33}(\omega) + A_{44}(\omega+\Omega)  \ A_{44}(\omega) \nonumber\\
  &+& \left. 2 A_{12}(\omega+\Omega)  \ A_{12}(\omega)+ 2 A_{34}(\omega+\Omega)  \ A_{34}(\omega) - 2 A_{14}(\omega+\Omega)  \ A_{14}(\omega) - 2 A_{23}(\omega+\Omega)  \ A_{23}(\omega)\right. \nonumber\\
  & - & \left. 4 A_{13}(\omega+\Omega) \ A_{13}(\omega)\right.
\nonumber\\
\label{lambda}
\end{eqnarray}

\end{document}